\documentclass[11pt,letterpaper]{article}
\usepackage{jcappub}
\usepackage{bbm,bm}
\usepackage{mathrsfs}
\usepackage{slashed}
\usepackage{caption}
\usepackage{epstopdf}
\usepackage[normalem]{ulem}
\usepackage[bottom]{footmisc}
\usepackage{subcaption}
\usepackage{bbold}
\usepackage{titlesec}
\usepackage{threeparttable}
\usepackage{booktabs}
\usepackage{changepage}
\usepackage[utf8]{inputenc}

\usepackage{grffile}

\usepackage{graphicx}  
\usepackage{dcolumn}   
\usepackage{bm}        
\usepackage{amssymb}   
\usepackage{setspace}
\usepackage{amsmath, amssymb, setspace}
\usepackage{array}
\usepackage{booktabs}
\usepackage{caption}
\usepackage{indentfirst}
\usepackage{float}
\usepackage{lmodern}
\usepackage{multirow}
\usepackage{soul}
\usepackage[normalem]{ulem}

\usepackage{braket}
\usepackage{comment}

\usepackage[draft]{pgf}

\newcommand{\figw}[2]{\includegraphics[width=#1\textwidth]{#2}}

\newcommand{\DoBox}[1]{\begin{center}
\color{red}\fbox{
\begin{minipage}{0.9\textwidth}

\end{minipage}}
\end{center}}

\newlength{\myimageoversize}
\newsavebox{\myimage}

\titleformat{\subsubsection}
 {\normalfont\fontsize{12}{17}\itshape}{\thesubsubsection}{1em}{}

\title{\huge{Upper limit on the diffuse extragalactic radio background from GZK photon observation}}

\author[a]{Graciela B. Gelmini,}
\author[b\, , c]{Oleg Kalashev,}
\author[d]{and Dmitri Semikoz}

\affiliation[a]{Department of Physics and Astronomy, University of California Los Angeles,\\
 Los Angeles, CA 90095-1547, USA}
 \affiliation[b]{Institute for Nuclear Research of the Russian Academy of Sciences, Moscow, 117312 Russia} 
  \affiliation[c]{Novosibirsk State University, Pirogova 2, Novosibirsk, 630090 Russia}
 \affiliation[d]{APC, Universit\`{e} Paris Diderot, CNRS/IN2P3, CEA/IRFU, Sorbonne Paris Cit\`{e},\\ 119 75205 Paris, France}

\makeatletter
\gdef\@fpheader{}
\makeatother
\begin{document}

\abstract{ 
Here we point out that an observation of  Ultra-High Energy Cosmic Ray (UHECR) photons, ``GZK photons", could provide an upper limit on the level of the Extra-Galactic Radio Background, depending on the level of UHECR proton primaries (to be determined after a few years of data taking by the Pierre Auger Observatory upgrade AugerPrime).
We also update our 2005 prediction of the range of GZK photon fluxes expected from proton primaries.
}

\maketitle

\section{Introduction}

The level of the extragalactic radio background (EGRB) has become a point of considerable interest in recent years. The  Absolute Radiometer for Cosmology, Astrophysics and Diffuse Emission (ARCADE-2) balloon-borne bolometer  detected in 2009~\cite{Fixsen:2009xn} an excess over the cosmic microwave background radiation (CMB) at high radio frequencies, 3 to 90 GHz,  after subtraction of the CMB and a model of Milky Way emission of more than double what is expected from known extragalactic point sources. Measurements from the Long Wavelength Array at 40 to 80 MHz and others~\cite{Dowell:2018mdb}, agree with the high ARCADE 2 level.

An overview of the state of the subject as of 2018~\cite{Singal:2017jlh}  showed that if the background seen by ARCADE-2 and other low-frequency experiments is  extragalactic, its origin is an important open question in astrophysics, since its sources would have to be very faint and much more numerous than known galaxies.  Besides the sources would need to be non-thermal due the spectral index of the background. The detected excess was found to be an isotropic component with  an antenna temperature as function of the frequency with a power law spectrum of index -2.58, flatter than known radio sources.  

Several exotic extragalactic  mechanisms have also been studied to explain the excess such as annihilating axion-like dark matter~\cite{Fraser:2018acy}  or dark photons~\cite{Pospelov:2018kdh}, supernova explosion of Population III stars~\cite{Jana:2018gqk}, superconducting cosmic strings~\cite{Brandenberger:2019lfm}, radiative decay of relic neutrinos to sterile neutrinos~\cite{Chianese:2018luo}, thermal emission from quark nugget dark matter~\cite{Lawson:2018qkc} and accreting astrophysical black holes~\cite{Ewall-Wice:2018bzf, Ewall-Wice:2019may}  or primordial black holes~\cite{Mittal:2021dpe}. 

A local origin of the mentioned background in the Local Bubble, a low-density cavity in the interstellar medium around the Solar system, has been proposed as an alternative~\cite{Krause:2021xav}.

Understanding if the mentioned radio background is of extragalactic or local origin is essential to study the expected signature in the 60 to 80 MHz range  associated with the 21 cm spin-flip transition of HI from the epoch of reionization. This is where EDGES (the Experiment to Detect the Global Epoch of Reionization Signature)~\cite{Bowman:2018yin}  has found an  absorption feature in 2018. Even a radio background with intensity 0.01 of the ARCADE-2~\cite{Fixsen:2009xn} result would have an observable effect at high-redshift 21 cm observations~\cite{Feng:2018rje, Jana:2018gqk}.

Here we explore the possibility of obtaining an upper limit on the EGRB if photons are detected in  Ultra-High Energy Cosmic Rays (UHECRs) given a measured fraction of proton primaries in UHECRs.
The origin of UHECRs, the cosmic rays with energies above $10^{18}$ eV, is a long standing  puzzle in astrophysics (for a recent review see e.g.~\cite{Kachelriess:2019oqu}). In particular, the composition (namely the fraction of primary protons and other nuclei) of the cosmic rays with energies beyond the Greisen-Zatsepin-Kuzmin (GZK) feature~\cite{Greisen:1966jv, Zatsepin:1966jv}  at $4\times 10^{19}$~eV has still large uncertainties (see e.g.~\cite{Kuznetsov:2020hso, Coleman:2022abf} and references therein). UHECRs are only observed through extensive air showers they produce in the atmosphere, what makes determining the nature  of their primary particles subject to the uncertainties of hadronic interaction models. The existing measurements have large errors and may contain unknown systematic effects. At EeV energies,  Pierre Auger Observatory (Auger)  and the Telescope Array (TA), find  a predominantly light composition with a large fraction of primary protons.  Above 2 EeV  Auger favors a mixed composition with mean mass steadily growing, while TA data seems to be in tension with this result, favoring a lighter composition~\cite{Bergman:2021djm}. However Auger and TA data are compatible within the current statistical and systematic uncertainties. At energies above the GZK suppression,  the total number of detected events is relatively small, thus the composition is even more uncertain~(see e.g.~\cite{PierreAuger:2017tlx, TelescopeArray:2018bep, Coleman:2022abf}). 

The results of Auger and TA indicate that UHECR primaries consist mostly of protons and heavier atomic nuclei, but there could be also primary photons.  Photons and neutrinos are produced by interactions of ultra-high energy charged particles with the background medium, either at the source site or during the propagation. Both Auger (since 2007~\cite{PierreAuger:2006lld}) and TA have searched for UHECR photon primaries and obtained upper limits on their flux~\cite{TelescopeArray:2020hey, PierreAuger:2021mjh, Rautenberg:2021vvt, Coleman:2022abf}. UHECR photons would be preferentially produced by proton  photoproduction of pions on the microwave background, in the decay of $\pi^0$, the process that leads to the GZK effect. These are ``GZK photons"~\cite{Gelmini:2005wu, Gelmini:2007sf,Gelmini:2007jy}. The observed flux suppression at $4\times 10^{19}$~eV could also be due to photon disintegration of heavy nuclei or a limit in the maximum particle energy reached at the sources, in which case many less ultrahigh-energy photons would be produced (e.g. considering only Fe primaries reduces the level of photons by more than an order of magnitude, see e.g. Fig.~7 of~\cite{Rautenberg:2021vvt}). The reason is that UHECR nuclei of atomic number A and energy E disintegrating at the GZK threshold produce protons, neutrons and lighter nuclei of energies $\simeq E/A$, which generally are below the photo-pion production threshold. Since in the following we will be relying on models that predict the maximum number of photons, we are going to consider only the production due to proton primaries.

The composition of UHECR primaries will be much better known after AugerPrime~\cite{PierreAuger:2016qzd, PierreAuger:2016qzd} (see also~\cite{Coleman:2022abf} and references therein)  operates for several years. This upgrade of Auger focuses on achieving mass-composition sensitivity for each extended air shower measured by its upgraded surface detector through multi-hybrid observations. If its planned resolution is achieved, AugerPrime should be able to distinguish between iron and proton primaries on an event-by-event basis at 90\% C.L. and even separate iron from the CNO group at better than 50\% CL. One of its design goals is to identify within five years of operation a proton fraction as low as 10\% with 5$\sigma$ statistical significance if such a component exists at the highest energies. AugerPrime is expected to be completed in 2023 and operate until at least 2032.

If it is confirmed that protons account for a certain fraction of the UHECRs, the level of UHECR photons will give an indication of the EGRB level, since these photons are absorbed in this background. In this paper we show the upper limits on the EGRB that an observation of UHECR photons would imply, given a certain fraction of UHECR proton primaries.  

The plan of the paper is the following. In Section 2, we explain how we model the sources and the propagation of particles. In Section 3, we present a revised version of our 2005 Fig.~17~\cite{Gelmini:2005wu} showing the range of integrated GZK photon  fluxes expected depending on the EGRB assumed.   In Section 4 we show our results on the upper bounds that an observation of UHECR photons could impose on the EGRB. In Section 5 we briefly conclude.

\section{Diffuse GZK photon flux calculation} \label{sec:calculations}

 To compute the flux of GZK photons produced by an homogeneous distribution of sources emitting originally only protons we use a numerical code~\cite{Kalashev:1999ma, Kalashev:2000af, Kalashev:2014xna}, which calculates the propagation of protons and photons using the kinematic equation approach and the standard dominant processes (see e.g.~\cite{Bhattacharjee:1999mup}).

The calculation of the predicted GZK photon fluxes requires assumptions about the characteristics of the UHECR sources and backgrounds.

\begin{figure}[!t]
    \centering
    \figw{.9}{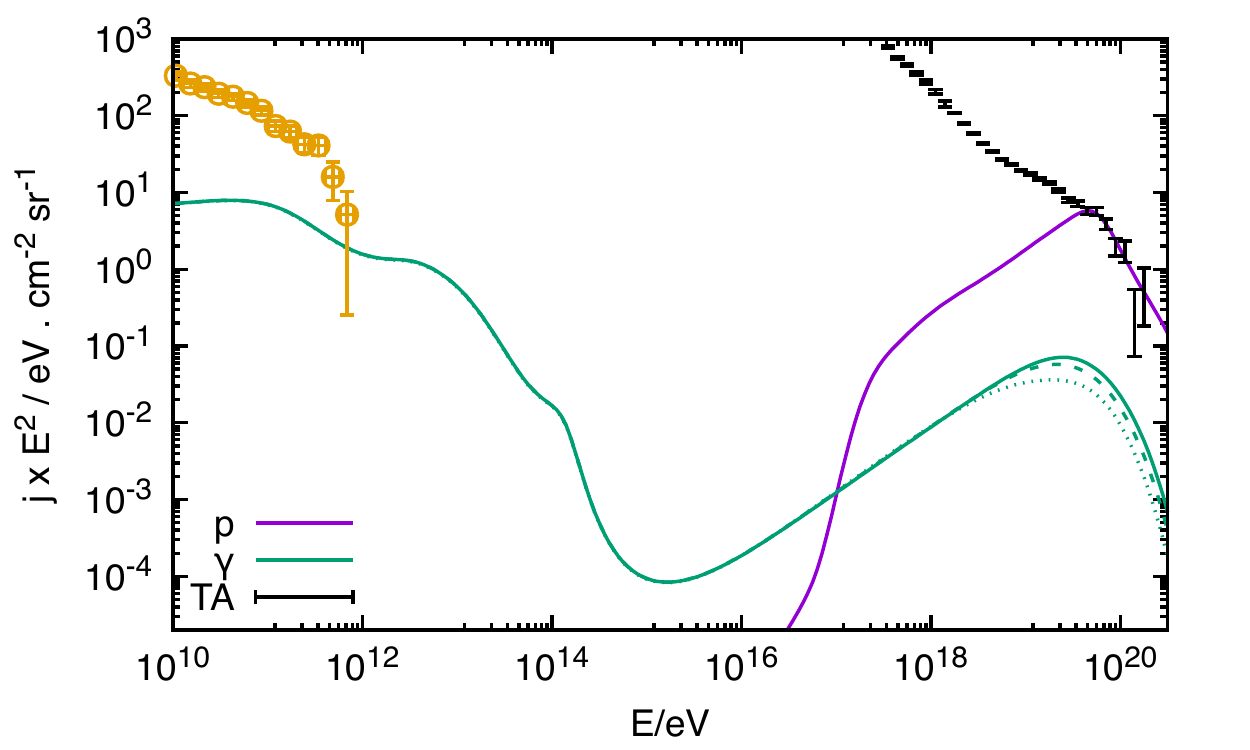}
    \caption{Example of the photon spectrum (green) produced by the proton (magenta) corresponding to
    the source model source model with $\alpha=0$, $E_{\max max}=10^{20}$, $m=0$, $z_{\rm max}=2$ and $z_{\rm min}=0$, 
    assuming that 100\% of the UHECR  TA spectrum\cite{Ivanov:2020rqn} above $4\times 10^{19}$ eV consist of protons. The green solid, dashed and dotted UHECR photon lines  corresponds to 10\% of the Clark et al.~\cite{Clark} EGRB and the higher and lower models of Protheroe and Biermann~\cite{Protheroe:1996si} respectively.  The radio backgrounds were scaled down by the factor $F_{\rm rad}= 0.1$ because Fig.~\ref{fig:spectra} shows that this is the level for which the predicted UHECR photon flux could soon be observed, namely the level of current Auger upper limits. The isotropic diffuse photon flux measured by Fermi LAT~\cite{Fermi-LAT:2014ryh} is shown in ochre. }
    \label{fig:spectra}
\end{figure}

Potential sources which could emit protons and nuclei with energy above  10$^{18}$ eV include gamma ray bursts, active galactic nuclei and starbursts galaxies. We adopt some  simplifying assumptions usually made about sources, namely that the UHECR emission rate is the same for all sources, that the spectrum and composition is independent of the redshift of the source, and that the emission rate is well described by a power-law spectrum.  The parameterization of the initial proton  flux we use is
\begin{equation}
F(E) = f~ \frac{1}{E^\alpha}~ \exp{\left(\dfrac{E}{E_{\rm max}}\right)}~.
\label{proton_flux}
\end{equation}
The power law index $\alpha$ and maximum energy $E_{\rm max}$ are considered free parameters. It is possible to adopt  $\alpha=0$ (as we do in Section 4), if the fraction of proton primaries is small and thus protons do not need to account for the UHECR spectrum. Otherwise, only the highest energy portion of this spectrum could be accounted for (see Fig.~\ref{fig:spectra}). The value $\alpha=2.7$ roughly corresponds to the slope of the UHECR spectrum. Larger values of  $\alpha$ disfavor having protons at higher energies.
The dependence of the GZK photon flux on  the maximum energy $E_{\rm max}$ is more significant as $\alpha$ decreases. Since we are interested in favoring the production of GZK photons, here we assume  $E_{\rm max} = 1 \times 10^{20}$~eV, a generous value considering that we assume a exponential cutoff (as opposed to a step function as we assumed in our earlier work~\cite{Gelmini:2005wu}).  The amplitude $f$ is fixed by requiring the final flux from all sources to be just below the observed UHECR spectrum, which we assumed to be that measured by TA\cite{Ivanov:2020rqn}.  The UHECR spectrum has been measured by Auger~\cite{PierreAuger:2020qqz} and TA~\cite{Ivanov:2021mkn}. A joint analysis showed that the TA spectrum is higher above the GZK cutoff~\cite{TelescopeArray:2021zox}. We use it in this paper because it favors the production of GZK photons.  
 
The evolution of the density of the sources can be parameterized as
\begin{equation} \label{source-evolution}
n(z) = n_0 (1+z)^{3+ m}~  \Theta (z_{\max}-z) \Theta (z-z_{\min}) \,,
\end{equation}
where $m$ is a real parameter and $z_{\min}$ and $z_{\max}$ are respectively the redshifts of the closest and most distant sources. In Section 4 we will  assume $m=0$, which  corresponds to non-evolving sources with constant density per comoving volume, since we are interested in estimating the maximum level of UHECR sources could produce. For the same reason we
assume $z_{\min}=0$. And we take $z_{\max}=2$, since sources with $z>2$ have a negligible contribution  to the UHECR flux  above $10^{18}$~eV.

UHECR protons could interact with extragalactic magnetic fields (EGMF) during their propagation, but this effect is negligible for fields below $10^{-11}$ G (see Fig.~5 of ~\cite{Gelmini:2005wu}), as found in constrained simulations.  Numerical cosmological simulations  give various predictions of  the EGMF 
strength~\cite{Garcia:2021cgu,Sigl:2004yk,Dolag:2004kp, Hackstein:2016pwa,Hackstein:2017pex}. Although observationaly  EGMF are bound to be between  $10^{-15}$ G~\cite{Neronov_2010, Taylor_2011}  and  $10^{-10}$ G~\cite{Neronov:2021xua}, constrained simulations, in which structure formation arguments are combined with measured fields in galaxy clusters, indicate that in the voids they should not exceed  $10^{-12}$ G~\cite{Dolag:2004kp}.  Thus extragalactic deflections could only happen close to the source or close to our galaxy if they are part of a filament~\cite{Courtois:2013yfa} with larger EGMF. Here we assume that EGMF are smaller than $10^{-11}$G in the voids and neglect their influence on GZK photons.

Photons with energies $E_\gamma >$ TeV pair produce electrons and positrons on background photons, mostly CMB photons except at  $E_\gamma > 10^{19}$ eV, initiating a cascade, which stops when the photons in it have $E_\gamma \simeq$ MeV to TeV. The flux of these extragalactic gamma ray background  has been measured by Fermi-LAT~\cite{Fermi-LAT:2014ryh} and is shown in Fig.~\ref{fig:spectra} (ochre points with their error bars). The main extragalactic point sources contribution to the Fermi data at high energies  are blazars. The contribution of unresolved blasars to the diffuse gamma-ray background  dominates at high energies ~\cite{Neronov:2011kg,DiMauro:2013zfa}.
The Fermi collaboration determined that up to 86\%  of the EGRB comes from unresolved blazars ~\cite{Fermi-LAT:2015otn}, what strongly constrains UHECR models that predict a large flux of UHECRs.

The main source of energy loss of photons with $E_\gamma > 10^{19}$ is pair production on the radio background and the uncertainty in this background translates into uncertainty in the photon energy-attenuation length.  As in our previous 
work~\cite{Gelmini:2005wu, Gelmini:2007sf,Gelmini:2007jy} we consider
three models for the EGRB: the background based on estimates by Clark et al.~\cite{Clark}  and the two models of Protheroe and Biermann~\cite{Protheroe:1996si},  both leading to a larger absorption of GZK photons than the first. The lowest model of Protheroe and Biermann  has a level of about 0.1 of the ARCADE-2~\cite{Fixsen:2009xn} measurement.

Fig.~\ref{fig:spectra} shows an example of the secondary photon spectra (green lines), assuming a UHECR  composition of only protons above the GZK cutoff,  with a source model $\alpha=0$, $E_{\rm max}= 1\times 10^{20}$, $m=0$, $z_{\rm max}=2$ and $z_{\rm min}=0$. The proton spectrum with amplitude $f$ such that its maximum is at the level of the  observed TA UHECR flux~\cite{Ivanov:2020rqn} (black data points) is shown in magenta. This model provides a good fit to the TA spectrum above the GZK cutoff. The three EGRB models just mentioned result in the three different secondary photon spectra shown above $10^{18}$ eV. The green solid, dashed and dotted UHECR photon lines  corresponds to 10\% of the Clark et al.~\cite{Clark} model,  10\% of the higher and 10\% of the lower models of Protheroe and Biermann~\cite{Protheroe:1996si}, respectively.  The radio backgrounds were scaled down by 10\%, i.e. bythe factor $F_{\rm rad}= 0.1$, because Fig.~\ref{fig:spectra} shows that this is the radio background level for which the predicted UHECR photon flux could soon be observed, namely for which this predicted flux is at the level of current UHECR photon experimental  upper limits.

\section{Expected range of the GZK photon flux}

\begin{figure}[htb]
\begin{center}
\includegraphics[width=.98\textwidth]{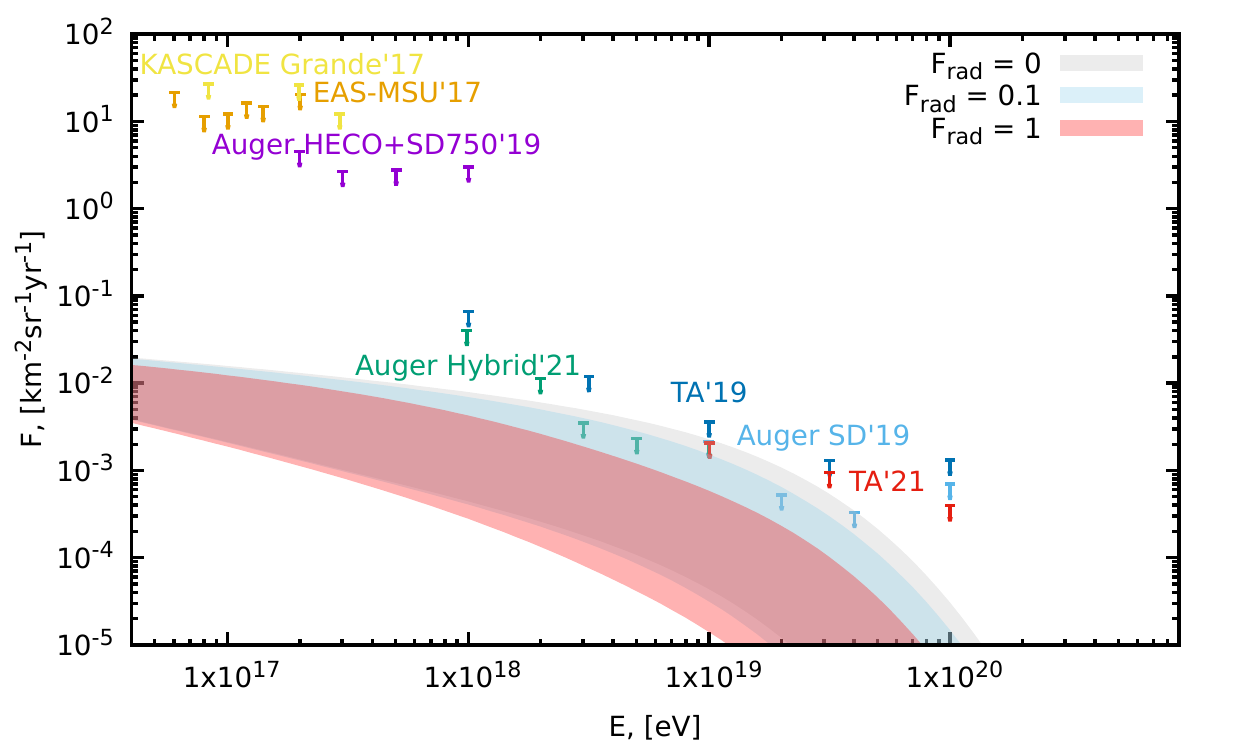}
\caption{Bands of predicted UHECR photon fluxes assuming only proton primaries (whose maximum is at the level of the TA UHECR spectrum~\cite{Ivanov:2020rqn}) for source parameters $0< \alpha < 2.8$, $-3 < m <3$ and  $E_{max}=10^{20}$~eV and the injection spectrum in Eq.~(\ref{proton_flux}). The bands in pink, sky blue and gray assume respectively the lower radio background model of Protheroe and Biermann~\cite{Protheroe:1996si} multiplied by the factors $F_{\rm rad}=1$, $F_{\rm rad}=0.1$ and $F_{\rm rad}=0$.  Also shown are past and present experimental upper limits by Auger~\cite{Rautenberg:2021vvt,PierreAuger:2021mjh}, TA~\cite{TelescopeArray:2018rbt,TelescopeArray:2021fpj}, KASCADE Grande~\cite{KASCADEGrande:2017vwf} and EAS-MSU~\cite{Fomin:2017ypo}.
\label{fig:potential}
}
\end{center}
\end{figure}

In this section we update our 2005 results presented in Fig.~17~\cite{Gelmini:2005wu} for the expected range of integrated GZK photon fluxes. In 2005 only very extreme source models could predict fluxes at the level of the experimental UHECR photon upper limits. These limits have become much better now~\cite{Rautenberg:2021vvt, PierreAuger:2021mjh, TelescopeArray:2018rbt,TelescopeArray:2021fpj}, as can be seen in Fig~\ref{fig:potential}.

Fig~\ref{fig:potential} shows the band of UHECR photon predicted fluxes assuming only proton primaries (whose maximum is at the level of the TA UHECR spectrum~\cite{Ivanov:2020rqn}) for a range of source parameters assuming the injection spectrum in Eq.~(\ref{proton_flux}): $0< \alpha < 2.8$, $-3 < m <3$ and, as throughout in this paper, $E_{\rm max}=1 \times 10^{20}$~eV. The three bands in pink, sky blue and gray assume respectively the lower radio background model of  Protheroe and Biermann~\cite{Protheroe:1996si} ($F_{\rm rad}=1$), 10\%  of it (i.e. the model mutiplied by the factor $F_{\rm rad}=0.1$) and no radio background (i.e. $F_{\rm rad}=0$).

The figure shows that the radio background level of 10\% of the lower 1996 Protheroe and Biermann model leads to predicted photon fluxes at the level of the current photon upper limits.  This EGRB level, 10\%  of the lower 1996 Protheroe and Biermann prediction, corresponds to 1\% of the ARCADE-2 level~\cite{Fixsen:2009xn} (a level predicted in some recent models~\cite{Feng:2018rje, Jana:2018gqk}).

\begin{figure}[!t]
    \centering
    \figw{.9}{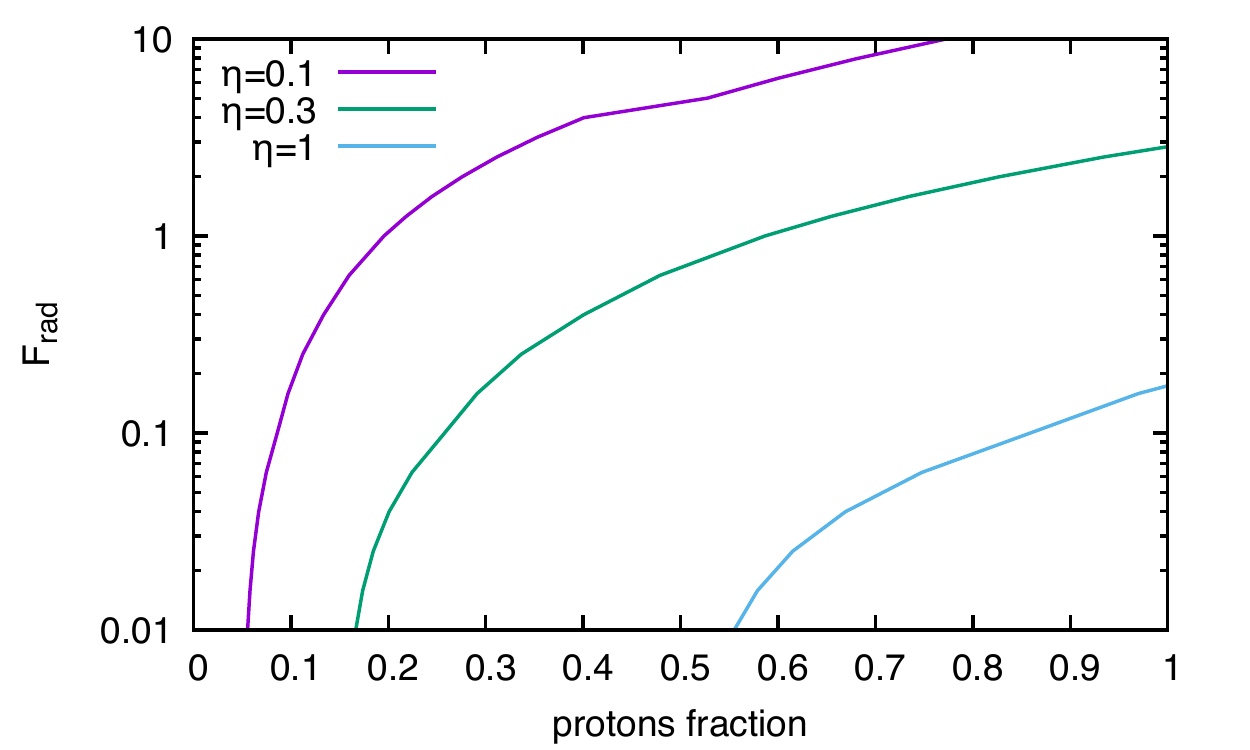}
    \caption{ 
    Upper limit on the EGRB, given as a factor $F_{\rm rad}$ times the lower radio background prediction of  Protheroe and Biermann~\cite{Protheroe:1996si},  which could be achieved if a GZK photon flux is detected at a  level  $\eta=0.1$ (magenta), $\eta=0.3$ (green),  and $\eta=1$ (blue) of the present Auger UHECR photon limits~\cite{Rautenberg:2021vvt,PierreAuger:2021mjh}, shown as function of the proton fraction in UHECRs (to be determined in the future by AugerPrime). The limits for other $\eta$ values can be easily obtained since they scale linearly with $\eta$. The production of photons from primaries other than protons is neglected here (which would not be consistent with very small proton fractions, close and below 0.1). The source model parameters assumed are $\alpha=0$, $E_{max}=10^{20}$, $m=0$, $z_{max}=2$, $z_{min}=0$. $F_{\rm rad}=10$ corresponds to the ARCADE-2 measurement~\cite{Fixsen:2009xn}. 
    }
    \label{fig:limits}
\end{figure}

\section{EGRB upper limit expected from UHECR photon observation} 

The limits on UHECR photons have improved so much over the years, as shown in Fig.~\ref{fig:potential}, that reasonable although optimistic production models yeald fluxes that could be detected in the near future. Also, as mentioned in the introduction, much better measurements of the proton fraction of UHECR primaries is expected to be obtained after about 5 years of operation of AugerPrime.

Here we compute the flux of UECR photons neglecting the production of photons by primaries other than protons,  which is not correct for proton fractions close and below 0.1. We assume a reasonable but optimistic model that favors the production of photons. Then, for each primary proton fraction, an observation of the UHECR photon flux would provide an upper limit on the EGRB, since a larger radio background would have suppressed the photon level to be below that observed. Namely if the radio background would be larger than the limit we find, even the maximum production model would lead to no signal.

For our optimistic production model we use the injection spectrum in Eq.~(\ref{proton_flux})  with parameters $\alpha=0$ and, as throughout in this paper, $E_{max}=10^{20}$~eV,  and the evolution parameters $m=0$, $z_{\rm max}=2$, $z_{\rm min}=0$.   It is appropriate to assume $m=0$,  which  corresponds to non-evolving sources with constant density per comoving volume, since the high energy photons come from close by, thus the effect of any evolution of sources is small. The choices $z_{\rm max}=2$, $z_{\rm min}=0$ are also reasonable, since sources with $z > 2$ have a negligible contribution to the UHECR flux above $10^{18}$ eV.

As it is clear from Fig.~\ref{fig:potential}, the most restrictive present limits on the GZK photon flux can be found in the energy range $18.5<$log(E/eV)$<19.3$ (the exact energy depend on the value of $F_{\rm rad}$).
Let us assume that in the future the UHE photon flux is detected at the level $\eta$ times the current limit at the most restrictive point. 
To obtain the minimal fraction of UHECR protons needed to generate enough GZK photons if $\eta=1$, we divide each present integrated photon flux limit by the predicted photon flux, computed  assuming a 100\% UHECR proton primary composition and a particular level of the radio background, and find the minimum value of this ratio among all the present limits. This minimum ratio determines the minimum fraction of protons in the UHECRs required to produce the photon signal. Clearly, since we consider only photon production from primary protons, for other values of $\eta$ the required  minimal proton fraction scales linearly with $\eta$. The constraints obtained in this way for different values of the EGRB scale factor $F_{\rm rad}$ are shown in Fig~\ref{fig:limits}.  Since we are assuming that the fraction of proton primaries will be measured by AugerPrime, we reverse the constraints, and show limits on $F_{\rm rad}$. 

In Fig~\ref{fig:limits} the upper limits on the EGRB are shown as function of the UHECR proton fraction to be determined by AugerPrime, in terms of the factor $F_{\rm rad}$ by which  the lower radio background prediction of  Protheroe and Biermann~\cite{Protheroe:1996si} should be multiplied, assuming an observation of UHECR photons at the level of the present limits or smaller by a factor $\eta$.  Only lines for $\eta=0.1$ (magenta), $\eta=0.3$ (green),  and $\eta=1$ (blue) are shown however the lines scale linearly with the fraction $\eta$ so the line for any other $\eta$ value can be easily obtained. Notice that $F_{\rm rad}=10$ corresponds to the ARCADE-2 measurement~\cite{Fixsen:2009xn}. Notice that for a particular value of $F_{\rm rad}$, the corresponding proton fraction for a particular $\eta$ is the minimum proton fraction compatible with the photon observation at the level indicated by $\eta$.

\section{Conclusions}
\label{sec:conclusions}

 In this paper we studied the possibility of deriving an upper limit on the diffuse extragalactic radio background from a possible future detection of the  GZK photon flux.
Our main result is summarised in Fig~\ref{fig:limits}.
Assuming the photon flux observed is at a level  of the factor $\eta$ times the current upper limits on UHECR photons, and given a particular proton fraction in the  UHECRs above 40 EeV one can establish an upper limit on the diffuse radio background. For example, for a photon flux three times lower than the present limits, i.e. with  $\eta=0.3$, and a proton fraction 0.5 (which is consistent with current TA data) one can exclude a radio background above the lower radio background prediction of  Protheroe and Biermann~\cite{Protheroe:1996si}.



\vspace{0.5cm}

\acknowledgments
The work of GBG  was  supported in part  by  the  U.S.  Department  of Energy (DOE) Grant No.  DE-SC0009937.   Work of O.K. is supported in the framework of the State project “Science” by the Ministry of Science and Higher Education of the Russian Federation under the contract 075-15-2020-778. We thank Peter L. Biermann for several clarifying discussions about radio background observations and models.


\bibliography{GZK2022}
\bibliographystyle{bibi}

\end{document}